\documentclass[preprints,article,accept,moreauthors,pdftex]{Definitions/mdpi}

\firstpage{1} 
\pubvolume{1}
\issuenum{1}
\articlenumber{0}
\pubyear{2021}
\copyrightyear{2020}
\datereceived{} 
\dateaccepted{} 
\datepublished{} 
\hreflink{https://doi.org/} 

\usepackage[utf8]{inputenc}
\usepackage[english]{babel}
\makeatletter\AtBeginDocument{\let\@elt\relax}\makeatother
\usepackage{amsmath}

\newcommand{\degg}[0]{^\circ}
\newcommand{\avg}[1]{\langle #1 \rangle}
\newcommand{\defeq}{\stackrel{\text{def}}{=}}

\Title{The violation of Bell-CHSH inequalities leads to different conclusions depending on the description used}

\TitleCitation{The violation of Bell-CHSH inequalities}

\Author{Aldo F.G. Solis-Labastida\orcidA{}$^{1}$,Melina Gastelum \orcidB{}$^{2}$ and Jorge G. Hirsch\orcidC{}$^{1,}$*}

\AuthorNames{Aldo F.G. Solis-Labastida, Melina Gastelum and Jorge G. Hirsch}

\AuthorCitation{Solis-Labastida, A.F.G.; Lastname, F.; Lastname, F.}

\address{%
$^{1}$ \quad Instituto de Ciencias Nucleares, Universidad Nacional Aut\'onoma de M\'exico,  C.P. 04510 CDMX., Mexico\\
$^{2}$ \quad Facultad de Filosof\'\i a y Letras, Universidad Nacional Aut\'onoma de M\'exico, C.P. 04510 CDMX, Mexico}

\corres{Correspondence: hirsch@nucleares.unam.mx}

\abstract{Since the experimental observation of the violation of the Bell-CHSH inequalities, much has been said about the non-local and contextual character of the underlying system. 
But the hypothesis from which Bell's inequalities are derived differ according to the probability space used to write them. 
The violation of Bell's inequalities can, alternatively, be explained assuming that the hidden variables do not exist at all, or that they exist but their values cannot be simultaneously assigned, or that the values can be assigned but joint probabilities cannot be properly defined, or that  averages taken in different contexts cannot be combined. All of the above are valid options, selected by different communities to provide support to their particular research program..}


\begin{document}
\section{Introduction}

Quantum mechanics is a probabilistic theory, due to the central role played by the Born rule to relate the calculations with the observations. These and other characteristics motivated Einstein and others to postulate that quantum mechanics must also be incomplete, statistical in its nature, and that some unknown, hidden variables must exist which could provide a deeper understanding and a more detailed description of the observations at the microscopic level\cite{einstein1935can,pais1979einstein}.  
Bell inequalities offer a way to contrast predictions of certain hidden variable theories with experimental observations \cite{bell1966problem}.  The straightforward argument, the simple mathematics involved and the strong conclusions drawn from it put it in a special place for the physicists, philosophers, and people interested in quantum phenomena. 

Since the first experimental observation of the violation of a Bell's inequality \cite{aspect1976proposed}, there have been a variety of interpretations with high impact in both the description of the world from a quantum physics perspective and in technological developments. For a large number of authors, models describing the quantum realm are not consistent with local realism  \cite{santos2016mathematical,bigbell2018}, local causality \cite{bell2004theory,maudlin2014what} and non-contextuality \cite{nieuwenhuizen2011contextuality}. A large variety of experiments have been performed to test the validity of the main assumptions needed to derive Bell's inequality \cite{kwiat1994proposal,gisin1999bell,barrett2002quantum,branciard2011detection}, and in 2015 the three most relevant loopholes were closed at the same time \cite{hensen2015loopholefree,Lynden2015,giustina2015}, which was interpreted as providing strong support to the impossibility to formulate local realistic theories that are compatible with the observed violation of Bell's inequalities.

In the area of quantum information, violations of Bell's inequality are used to characterize properties associated exclusively with quantum systems \cite{salavrakos2017bell}.  It has been postulated that the outputs of quantum measurements, when the system is prepared and measured in different basis, are ''intrinsically random'' \cite{zeilinger2005,Bera2017,grangier2018}, and guarantee the randomness of random number generators \cite{pironio2010randoma,colbeck2011private,fehr2013security}. 

Randomness and probability are closely related in the analysis of quantum phenomena \cite{khrennikov2019}. It has been argued that  the Hilbert space formalism of quantum mechanics is a new theory of probability \cite{pitowsky2006quantum}. While some authors have pointed out that non-locality, as well as rejection of realism, are only sufficient (but non-necessary) conditions for violation of Bell’s inequality \cite{khrennikov2009,khrennikov2020two} and that the Bell inequalities only need to be satisfied if all observables can be measured jointly \cite{demuynck1984alternative}, it seems that the analysis of the violation of Bell's inequality from the probability theory point of view is not fully understood in the physics community.  In this contribution our aim is to add more elements, in the common language of probability, to include a variety of interpretations of the violation of Bell's inequality beyond non-locality.

Our starting point is the probability space, which is presented in detail in the next section. We adhere to the view that different probability spaces lead to different Bell's inequalities \cite{jarrett1984physical, berkovitz2012world}. While the inequalities look similar, they are not equivalent. They are different mathematical objects with different sets of hypotheses which, when questioned, generate completely different conclusions: non-locality, contextuality (setting dependence), impossibility to assign values to unmeasured quantities, impossibility to define a joint probability distribution, etc. Our intention is to expose that there is no unique way to decide which inequality is employed in the analysis of a Bell type experiment, and the interpretation depends on this choice, which in turn elicits the ontological and epistemological possibilities according to each decision. 

Bell's inequalities refer to the probabilities of finding correlations between experimental outputs of a relatively simple experimental setup. Although the quantum mechanical description perfectly matches the experimental results, Bell's inequalities have consequences far beyond quantum physics, and they apply to a variety of generalized probabilistic theories \cite{janotta2014}. This adds to its relevance, but at the same time, there is no common basis to which all communities adhere for its application. We hope this contribution will serve as an invitation to recognize the diversity of interpretations, all of them equally valid in their own terms and with different consequences.

The article is organized as follows: Section II presents a short review of Kolmogorov axioms, Section III describes a Bell type experiment and the different probability spaces which can be employed in its description. Section IV contains the derivation of the different Bell's inequalities, describing the hypothesis involved in each of them. The interpretation of the violation of the inequalities are presented in  Section V, and the Conclusions in Section VI.

\section{A short review of Kolmogorov axioms}

In the measure theoretic view,  axioms of probability assume the existence of a sample space, an event space and a probability measure \cite{rosenthal2006first, kolmogorov2018foundations}. These three elements, commonly referred as the \textit{probability space}, must be properly defined when probability is invoked. 

One important motivation for this contribution is that, in many cases, operative rules are provided without making explicit the probability space in which they are being employed. In the case of physics, many textbooks, basic or advanced, particularly those devoted to quantum mechanics, tend to introduce probability as it were another physical quantity like charge, mass, etc. \cite{townsend2000modern, cohen-tannoudji1991quantum, sakurai2017modern, nielsen2010quantum}. We show in what follows that the implications of the violation of the Bell inequalities are radically affected when different probability spaces are employed.  

The sections below contain a modern presentation of the probability space \cite{gut2005probability, rosenthal2006first, chung2001course}.

\subsection{The probability space}

\subsubsection{The sample space}

The first element in the list is the \emph{sample space}, commonly represented by $\Omega$. This is a set whose elements represent the possible outputs of a trial or experiment. In a coin flip, the set: $\{heads,tails\}$ is the common election for a sample space.  In a dice the sample space are the numbers one through six,$\{1,2,3,4,5,6\}$. 

The sample space has to fully characterize the outputs of the experiment. That is its main feature. Every output not taken into account in the sample space will be ignored. For instance, in the coin flip case, the case where the coin lands on its edge is excluded. 

Probability is commonly discussed in relation with games like dices, coins, the roulette, etc. In those scenarios the sample space appears almost trivial, it is just the list of possible outcomes in the experiment. However, the situations where probability is applied can be far wider than them, making more challenging to define the sample space.

\subsubsection{The event space}

The second element in the list is the {\em event space}. This space is meant to represent the situations at which a probability is associated. It is composed of subsets of sample space. Each of these subsets is referred as an \emph{event}. While the sample space represents all the possible outputs, the event space can include the representation of more complex situations. 

For instance, in the dice example, the probability of the output being even is represented by the event $\{2,4,6\},$
i.e. the set that contains all the possible outcomes that fulfills the criterion "being even". Another example is "the output is six", the event is:
$\{6\},$ this is a set whose only element is $6$. Important events are $\{1,2,3,4,5,6\},$
which can be read as "any of the possible outputs", and  the "null event", that is the empty set $\{\}$ which can be interpreted as "none of the possible outputs". 

The event  space doesn't necessarily include all possible subsets of the sample space, it only needs to fulfill three conditions.

\begin{enumerate}
\item It must contain the sample space. Therefore there is an event, the "total" event, that contains all the possible outputs of the experiment. Just as we wrote above, for a dice this event is $\{1,2,3,4,5,6\}$.
 
\item If an event belongs to the space, then its "complement" also belongs to the space. For instance, if the total event is in the space, then the null event must also be in the space. For the dice, if the event containing all the even outputs is in the event space, then the event $\{1,3,5\} $, the one containing all odd outputs,
must also be in the space. 

\item The space must be closed under countable unions and intersections. That is, if the space has the event $\{2,4,6\}$
and also has the event $\{1,2,3,4\}$, then the event $\{1,2,3,4,6\}$ and the event  $\{2,4\}$
must also be in the space.
This can be interpreted as "the event where the output is even and is less than 5". 

In this example the word "and" refers to the intersection of subsets. In a similar way, the union of subsets is referred to with the word "or", as in "the event where the output is even or less than 5". 
\end{enumerate}

It is relevant to mention two extreme cases. An event space can be formed using only the "total event" and the "null" event. These two events are enough to have a proper event space. This is called the "smallest" event space possible. 

In the other side of the spectrum there is the "biggest" event space. This is the set of all the possible subsets, the power set of the sample space. In some cases it is a natural way to represent all the possible events, but in many situations there are restrictions which exclude some elements of the biggest event space.  

\subsubsection{The probability measure}

The third element in the list is the \textit{probability measure} $P$. It is a function that assigns a probability to each event, i.e. a real number. It is important to note that such function must be total, i.e. the probability measure must be defined in every element of the event space.

In the smallest event space, only the total event and the null event will have a probability. In the biggest event space, all the possible events have a probability. 

Having defined the probability space, the Kolmogorov axioms can be enunciated.

\subsection{Kolmogorov axioms}

The Kolmogorov axioms are considered the foundations of modern probability theory \cite{kolmogorov2018foundations}. Given a sample space $\Omega$, an event space $E$ and a probability measure $P$, such measure must fulfill:
\begin{align}
P(E) &\geq 0 ,\\
P(\Omega)& = 1 ,\\
	\text{if } E_i \cap E_j = \emptyset, & \text{ then  } &\nonumber \\
	P(E_1 \cup E_2 \cup \ldots) &= P(E_1) + P(E_2) + \ldots
\label{kolmogorovAxioms}
\end{align}
where $E,E_i,E_j$ are elements of the event space. This axioms constrain probabilities to have values between zero and one, and other common properties attributed to probabilities.

Notice that the third axiom refers to mutually exclusive events, those which cannot occur at the same time or in the same run \cite{gillies2000philosophical}.
It can be extended to include the case $E_1 \cap E_2 \neq \emptyset$ as
$$
P(E_1 \cup E_2) = P(E_1) + P(E_2) - P(E_1 \cap E_2).
$$
In what follows, we employ the notation $P(E_1, E_2) \equiv P(E_1 \cap E_2)$.

\subsubsection{Conditional probabilities}

One of the most important concepts in probability theory is the {\em conditional probability} $P(E_1|E_2)$ \cite{ross2018first}, the probability of event $E_1$ given that the event $E_2$ has occurred. In Kolmogorov's framework it is introduced as a derived concept
$$
P(E_1|E_2) \equiv P(E_1,E_2) / P(E_2).
$$

In the case two events are statistically independent, i.e. $P(E_1,E_2) = P(E_1)P(E_2)$, the conditional probability is unaffected by the introduction of the conditional, since $P(E_1|E_2) = P(E_1)$.

For the present analysis, it is important to highlight the fact that a conditional probability is not associated with an event in the Kolmogorov's framework. 

\subsubsection{Probabilities and relative frequencies}

Relative frequencies are the quantities actually observed in a experiment. When an experiment is repeated $M$ times, and an event takes place $N_E$ times, the {\em relative frequency} of the event $E$ is defined as $\frac{N_E}{M}$.

These quantities are expected to be approximately equal to probability values as $M$ grows:
\begin{equation}
	P(E) \approx \frac{N_X}{M} 
\end{equation}
How tight this equality should hold is not specified in the Kolmogorov's framework.

\subsubsection{Expected values}
\label{expectedValues}

In many situations it is useful to associate probability with variables instead of with events. To do so, it is a common practice to use {\em random variables} as an abbreviation for events. For instance, the probability $P(X=x)$ is the probability of the event "the quantity $X$ assumed the value $x$".

For any random variable in a probability space, with probability $P(X)$,  its expectation value is defined as $\avg{f(X)} \defeq \sum_i f(x) P(X=x)$. This quantity is expected to have a value close to the average value observed in experiments:
\begin{equation}
	\avg{X} \defeq \sum_{x} x P(X=x) \approx \sum_{i=1}^{M} \frac{x_i}{M} ,
\end{equation}
where $x_i$ is the result of the $i$-th run of the experiment.

Please note that the above definition does not specify the probability space to which these probabilities belong. For instance, $X$ can be in a probability space that also includes $Y$. In this case the definition of average is:
\begin{equation}
	\avg{X} \defeq \sum_{x,y} x P(X=x,Y=y) ,
\end{equation}
where $P(X=x,Y=y)$ is the joint probability, which allows to obtain $P(X=x)$ via marginalization:
$$
	P(X=x) = \sum_{y} P(X=x,Y=y).
$$

An important case for our purposes is that of binary variables, which can only assume values $0$ and $1$. Their expected values have the following property:
$$
	\avg{X} \defeq 0 \times P(X=0) + 1 \times P(X=1) = P(X=1)
$$

If there are other binary variables in the probability space, it is possible to extend the previous identity.
The product $XY$ is 1 if and only if X = Y = 1, and 0 otherwise. In this case, each expected value is equivalent to the probability of having the value 1
\begin{align}
	\avg{XY} &\defeq \sum_{x,y = \{0,1\}} x \,y  \,P(X=x,Y=y)  \\ \nonumber 
	&= (0 \times 0 ) P(X=0,Y=0) + 
	(0  \times 1) P(X=1,Y=0)  \\ \nonumber
	& \quad + (1 \times 0) P(X=0,Y=1) +
	 (1 \times 1) P(X=1,Y=1) \\ \nonumber
	&= P(X=1,Y=1) = P(XY=1)
	\label{xyproduct}
\end{align}

\section{Probabilities and Bell inequalities}

\subsection{Bell type experiment}

\begin{figure}[htp]
\centering
\includegraphics[width=0.7\textwidth]{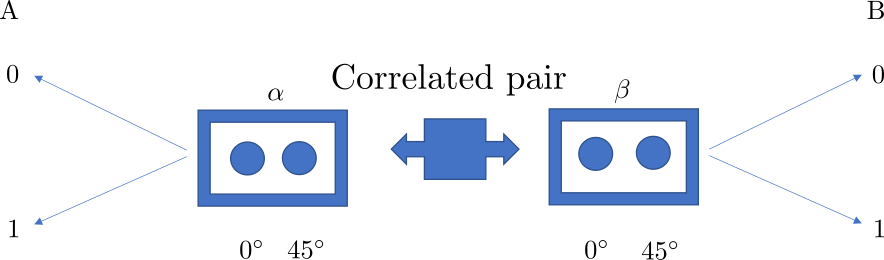}
\caption{Bell type experiment diagram. There are four quantities of interest in this setup: the~angles $\alpha, \beta$ in which the measurement is performed and the results $A, B$ of the measurement in each side.}
\label{figuraBell}
\end{figure}

A Bell type experiment \cite{aspect1976proposed, clauser1969proposed,genovese2005research} is schematically represented in Fig. \ref{figuraBell}. In each run of the experiment, a pair of entangled photons, particles, or other quantum systems, is generated, one photon traveling to the left and the other to the right. While the photons are in flight, a random selection is made in each side selecting an angle, which sets the polarization basis to be employed in the measurement, between two previously defined options. For simplicity of notation we employ here the pair $0\degg$ and $45\degg$ on both sides. Changing the angles to, for example, $22.5\degg$ and $-22.5\degg$ in one side, has no effect in the following discussion. 

Once the polarization basis is  selected, each photon travels through a polarized beam splitter with a single photon detector in each arm. The detector activated in each side is univocally associated with a polarization projection in the selected basis, allowing to assign a value to the polarization, in that basis, of each photon detected. To take into account the detector efficiency and presence of noise, the valid measurements are selected as those where one photon is detected in each side. 
In this way, all events include in each side one and only one angle selected and one and only one detector which registered a photon. 

After each run of the experiment there are four quantities registered: one angle selected and one detector activated in each side.

There are different probability spaces which can be employed to describe the previous experimental situation, and the Bell inequalities can be built on each one of them. Their details are given below.

\subsection{The probability space 1}

\subsubsection{The sample space 1}

In this space the possible results of the experiment are characterized by four variables, one angle selected and one detector activated in each side. A hidden variable $\lambda$ is introduced, which cannot be observed.
The sample space can be represented with quintuples of the following form:
\begin{align}
  (A,\alpha,B,\beta,\lambda).
\end{align}
Here, $A,B$ represent which detector is activated in each side. These variables can only have one of two values denoted $0$ and $1$. The other variables, $\alpha,\beta$,  represent the chosen orientation values. They have also only two options, represented by $0\degg$ and $45\degg$. A hidden variable $\lambda$ is also considered.
The sample space contains sixteen different quintuplets for a given value of $\lambda$: 
\begin{align}
\bigg\{
	&(A=0,\alpha=0\degg,B=0,\beta=0\degg,\lambda=\lambda'), 
	\nonumber \\
	&(A=0,\alpha=0\degg,B=0,\beta=45\degg,\lambda=\lambda'), 
	\nonumber \\
	&(A=0,\alpha=0\degg,B=1,\beta=0\degg,\lambda=\lambda'), 
	\nonumber \\
	&\hspace{3cm}\vdots
	\nonumber \\
	&(A=1,\alpha=45\degg,B=1,\beta=0\degg,\lambda=\lambda'), 
	\nonumber \\
	&(A=1,\alpha=45\degg,B=1,\beta=45\degg,\lambda=\lambda'),\ldots
\bigg\}
\label{sampleSpace1}
\end{align}

They represent all the possible combinations for the values of the variables. 
In a trial of the experiment, it is possible to know the value of the first four variables, however, the last one remains unknown. All possible values of $\lambda$ describe the same output of the experiment. 

\subsubsection{The event space 1}

Implicit in most uses of probability, it is common to choose as the event space the biggest event space, which includes all possible subsets of the sample space.
For instance, if side A sets an angle $\alpha=0\degg$ and gets a result $A=1$, the corresponding event is:

\begin{align}
	\bigg\{  
		(&A=1,\alpha=0\degg,B=0,\beta=0\degg,\lambda=\lambda'),\nonumber \\ 
		(&A=1,\alpha=0\degg,B=1,\beta=0\degg,\lambda=\lambda'), \nonumber \\
		(&A=1,\alpha=0\degg,B=0,\beta=45\degg,\lambda=\lambda'), \nonumber \\
		(&A=1,\alpha=0\degg,B=1,\beta=45\degg,\lambda=\lambda'),\ldots
	\bigg\},
\end{align}
i.e, all the quintuples that meet the requirement $A=1$ and $\alpha=0\degg$. This set contains these quintuples for each value of $\lambda$. 

\subsubsection{The probability measure 1}

For the third element of the probability space, the probability measure, it is enough to assume that it fulfills the previous requirements and Kolmogorov axioms. 
It assigns a real number between zero and one to each event in the event space 1.

For instance, the probability of the above mentioned event, side A sets an angle $\alpha=0\degg$ and gets a result $A=1$, is 
$$
	P(A=1,\alpha=0\degg).
$$

This probability is different to the probability that side A gets a result $A=1$  given the polarization is measured setting the angle at $\alpha=0$
$$
	P(A=1|\alpha=0\degg),
$$

This is a conditional probability, obtained evaluating  
$$
	P(A=1|\alpha=0\degg) = P(A=1,\alpha=0\degg) / P(\alpha=0\degg),
$$

Which one of them must be employed in the formulation of Bell inequalities will be explained in the next section.

\subsection{The probability space 2}

\subsubsection{The sample space 2}

A different sample space can be built, if it is assumed that in each run of the experiment, it is possible to assign values to the outputs of both orientations of the polarization. It is a picture closer to the classical view where, in a destructive measurement which only allows to measure one property of a system, out of many, in each procedure, the other properties are assumed to have definite values. 
In each run of the experiment, there is one and only one measured value of $A$, and one of $\alpha$ (and the same happens for $B$ and $\beta$).

The two possible outputs on the left side are denoted $A_{0\degg}$ and $A_{45\degg}$. They are assumed to have, in each run, values like $A_{0\degg}=1$ and $A_{45\degg}=1$. It represents a situation where, if the measurement is performed in the basis $0\degg$, the result will be $1$, and if it is performed in the basis at $45\degg$ the result will be also $1$. 

This sample space has also four quantities of interest: for side A, the results when the apparatus is set at $0\degg$ and set at $45\degg$. For side B there also are results for $0\degg$ and for $45\degg$. A hidden variable $\lambda$ is also added.

The sample space 2 can be written as:
\begin{align}
	\bigg\{
		&(A_{0\degg}=0,A_{45\degg}=0,B_{0\degg}=0,B_{45\degg}=0,\lambda=\lambda'),
		\nonumber \\
		&(A_{0\degg}=0,A_{45\degg}=0,B_{0\degg}=0,B_{45\degg}=1,\lambda=\lambda'),
		\nonumber \\
		&(A_{0\degg}=0,A_{45\degg}=0,B_{0\degg}=1,B_{45\degg}=1,\lambda=\lambda'),
		\nonumber \\
		&\hspace{3cm}\vdots
		\nonumber \\
		&(A_{0\degg}=1,A_{45\degg}=1,B_{0\degg}=1,B_{45\degg}=1,\lambda=\lambda'), \ldots
	\bigg\}.
\end{align} 
where all these quintuplets repeat for each value of $\lambda$.

\subsubsection{The event space 2 and the probability measure 2}

Again the chosen space for event space 2  is the biggest event space associated with sample space 2. Every event in it has an associated probability. These three elements form probability space 2. The probability measure is assumed to satisfy Kolmogorov axioms, i.e. the probability measure assigns values between $0$ and $1$ to all events.

\subsection{Events and probabilities in the analysis of the Bell experiments}

Sample spaces 1 and 2 have  similarities and differences which are worth to analyze in detail.

\subsubsection{The same event in the two probability spaces}

The same experimental outcome is associated with different events in each space. As an example, if the result is $0$ in the side A of the experiment with the angle $0\degg$, while in side B the result is $1$ with the experiment set up at angle $45\degg$, this event in event space 1 is:
\begin{align}
	\bigg\{	(A=0, \alpha=0\degg, B=1,& \beta=45\degg,\lambda=\lambda'),\ldots
	\bigg\}
\end{align}
where there is one quintuple for each value of $\lambda$.

In the event space 2 the set of quintuples compatible with these same conditions is:
\begin{align}
	\bigg\{
		&(A_{0\degg}=0,A_{45\degg}=0,B_{0\degg}=0,B_{45\degg}=1,\lambda=\lambda'),    \nonumber \\
		&(A_{0\degg}=0,A_{45\degg}=0,B_{0\degg}=1,B_{45\degg}=1,\lambda=\lambda'),    \nonumber \\
		&(A_{0\degg}=0,A_{45\degg}=1,B_{0\degg}=0,B_{45\degg}=1,\lambda=\lambda'),    \nonumber \\
		&(A_{0\degg}=0,A_{45\degg}=1,B_{0\degg}=1,B_{45\degg}=1,\lambda=\lambda'),\ldots
	\bigg\}
\end{align}
where these four quintuplets repeat for each value of $\lambda$.

This example shows that the same experimental situation has different events associated in the different event spaces. 
This can be seen in the following simple example, where only events associated with the left arm of the experiment are considered.

\subsubsection{The same probability in the two probability spaces}
 \label{theSameProbabilityIn}

In space 2, the probability that $A_{0\degg}=1$ is $P(A_{0\degg}=1)$. What is the probability of this event in space 1? 
It refers to those cases where side A gets the result $A=1$, but only for $\alpha=0\degg$. Note that the election 
$P(A=1,\alpha=0\degg)$ is not the right one, because, if $\alpha=0\degg$ is a very unlikely event, the previous probability will be very low due to the fact that the measurement apparatus in side A almost all the time is set in $\alpha=45\degg$. Therefore it is important to take into account only those cases in which $\alpha=0\degg$. This is represented by the conditional probability 
$$
P(A=1|\alpha=0) = \frac {P(A=1,\alpha=0\degg)} {P(\alpha=0\degg)}.
$$

It is worth to extend the previous argument a little further using an example. 
In the left hand side of Table \ref{experimentalTableExample}  the events in space 1 representing the output of each run are listed.
In the probability space 2 there is a value for $A_{0\degg}$ and $A_{45\degg}$ in every experiment, but it is only possible to measure one of them in each run of the experiment.  In the right hand side of Table \ref{experimentalTableExample}, both the observed and the unobserved outputs associated to each event in event space 2 are shown, with a circle enclosing the observed quantities. The value which was not measured could have been $0$ or $1$, but in each run it is assumed to have a value.

\begin{table}
\begin{center}
\begin{tabular}{|c|c|c|}
\multicolumn{3}{c}{space 1}\\
\hline
Run & $A_{}$ & $\alpha$ \\
\hline
{\bf 1} &~ {\bf 1} ~ & ~ ${\mathbf 0\degg}$ ~\\
2 & 1 & $45\degg$ \\
3 & 1 & $45\degg$ \\
{\bf 4} &~ {\bf 1} ~ & ~ ${\mathbf 0\degg}$ ~\\
5 & 0 & $0\degg$ \\
6 & 1 & $45\degg$ \\
\multicolumn{3}{c}{$\vdots$}
\end{tabular}
\qquad
\begin{tabular}{|c|c|c|}
\multicolumn{3}{c}{space 2}\\
\hline
Run & $A_{0\degg}$ & $A_{45\degg}$ \\
\hline
{\bf 1 }& \textcircled{\bf 1} & 0 \\
2 & 0 & \textcircled{1} \\
{\bf 3} & {\bf 1} & \textcircled{1} \\
{\bf 4 }& \textcircled{\bf 1} & 0 \\
5 & \textcircled{0} & 1 \\
{\bf 6} & {\bf 1} & \textcircled{1} \\
\multicolumn{3}{c}{$\vdots$}
\end{tabular}
\end{center}
\caption{An experimental run can be depicted in different probability spaces. On each line, the table on the left shows the experimental result depicted in probability space 1, and those on the right refers to the same experimental result in sample space 2.}
\label{experimentalTableExample}
\end{table}

In space 1, the number of events in which $A=1$ and $\alpha=0\degg$ is 2, and the total number of events is 6. It follows that 
$$P(A=1,\alpha=0\degg) = 2/6 = 1/3.$$

In space 2 the values of $A$ for $\alpha=0\degg$ are assumed to exist, even when the angle selected was $\alpha=45\degg$. In Table \ref{experimentalTableExample} both the measured and the hidden values are listed. They show $A_{0\degg} = 1$ in 4 of the 6 events. So, in space 2
$$P(A_{0\degg}) = 4/6 = 2/3.$$

Returning to space 1, $P(A_{0\degg}) = 2/3$ can be expressed as  the number of events in which $A=1$ given $\alpha=0\degg$, which is the conditional probability
$$
	P(A=1|\alpha=0)=\frac{P(A=1,\alpha=0)}{P(\alpha=0)}= \frac{1/3}{1/2} = 2/3.
$$
It is worth to highlight that the conditional probability is not associated with an event in space 1. It is a mere convention meant to represent the quotient of probabilities that do have an associated event.
  
The above discussion confirms the relation
\begin{align}
	P(A_{0\degg}=1) = P(A=1|\alpha=0\degg),
	\label{jointProbabilityA}
\end{align}
where the probability in the left hand side is in probability space 2 while the one in the right hand side is in probability space 1.
This equivalence is relevant in the analysis of Bell inequalities.

\subsubsection{Some events exist only in one probability space}

There are situations in which an event in one event space cannot be represented in the other. 
For instance, the event $\{\alpha=0\}$ in space 1 cannot be expressed in space 2, since there is no variable related to the chosen angle.

On the other hand, the event  $\{A_{0\degg}=0,A_{45\degg}=0\}$ exists in space 2, but in space 1 it has no meaning, because it would imply to select both polarization angles at the same time. This is not possible since it is not included in sample space 1.

\section{Two Bell-CHSH inequalities}

Bell-CHSH inequalities refers to a family of inequalities. One very common and useful is the CHSH inequality \cite{clauser1969proposed}.
\begin{align}
	-2 \leq \avg{XY} + \avg{X'Y} + \avg{XY'} - \avg{X'Y'} \leq 2
\label{chshInequality}
\end{align}
where the random variables $X,X',Y,Y'$ can take only the values $+1$ or $-1$. Please note that the previous expression leaves unspoken the probability space in which the averages take place.

It is useful to rewrite the above expression in terms of random variables $A,A',B,B'$ which take the values $+1$ or $0$ instead. In this form it reads
\begin{align}
	-1 \leq \avg{AB} + \avg{A'B} + \avg{AB'} - \avg{A'B'} - \avg{A} -\avg{B} \leq 0.
\label{chshInequality1}	
\end{align}
this will be the one used along this paper. 

As it was explained in section \ref{expectedValues}, employing  variables which can only take the values $1$ or $0$ allows to write their expected values as probabilities. For instance, $\avg{AB}$ is equal to the probability of having a value one in side A for $\alpha$ and also in side B for $\beta$. The relevant point here is that there are two probability spaces in which such probability can be written. As explained in section \ref{theSameProbabilityIn}, in probability space 1 the equality
\begin{align}
	\avg{AB} = P(A=1,B=1|\alpha,\beta) 
\end{align}
should hold. In contrast, in probability space 2 the equality
\begin{align}
	\avg{AB} = P(A_\alpha = 1, B_\beta = 1) 
\end{align} 
holds. Since every term in equation (\ref{chshInequality1}) can be expressed in either of the spaces, we end up with two possible inequalities. 

As the two probabilities have been proven equal, the matter of choosing a probability space to write the inequality appears irrelevant at first glance. However, both options will have completely different consequences when the violation of Bell-CHSH inequalities is analyzed.

To continue, a derivation of the inequality in the different spaces is exposed. Here we reproduce a derivation that simplifies the mathematical expressions \cite{berkovitz2012world} and enables to clearly state the mathematical hypotheses used. It is based in the following mathematical inequality. Given four real numbers $ a,a',b,b' \in [0,1]$, as shown in the Appendix, the following inequality holds
\begin{equation}
    -1 \leq ab + a'b + ab' - a'b' - a - b \leq 0
    \label{ineq}
\end{equation}

\subsection{Bell inequality in probability space 1}

The version of the inequality  (\ref{chshInequality1}) expressed in probability space 1 is the most commonly found in the literature, and the closest with the original intention of the inequality written by Bell \cite{bell1966problem,bell2004introduction,clauser1969proposed}. 

Since $\lambda$ is assumed to be unobserved in the experiments, its relative frequencies cannot be measured. If the inequalities are to be compared with experiment, the involved probabilities should not include $\lambda$. This is accomplished by averaging over the $\lambda$ variable. 

Based on the Kolmogorov's framework, the averaging is given by:
$$
P(A=1,B=1|\alpha,\beta)  =  \int_{\Lambda} P(A=1,B=1|\alpha,\beta,\lambda) P(\lambda|\alpha,\beta) \,d\lambda
$$
for the four possible values of $\alpha$ and $\beta$.

Two hypothesis are employed to manipulate the above conditional probabilities. 
One is the \textit{locality hypothesis} \cite{bell2004theory}:
\begin{align}
	P(A,B|\alpha,\beta,\lambda) = 
	P(A|\alpha,\lambda) P(B|\beta,\lambda).
\label{localityHypothesis}
\end{align}
Based on the concept of statistical independence, it is commonly interpreted as the requirement that the polarization angle $\alpha$ selected in the left hand side does not influence the value of the output $B$ on the right hand side, and vice versa

As there are four different settings, one for each pairs of angles $\alpha,\beta$, it is necessary to add  the hypothesis of $\lambda$-independence, sometimes referred to as settings independence \cite{ciepielewski2020superdeterministic}:
\begin{align}
	P(\lambda|\alpha,\beta)=P(\lambda).
\label{lambdaIndependence}
\end{align}
They allow to write
$$
P(A=1,B=1|\alpha,\beta)  =  \int_{\Lambda} P(A=1|\alpha,\lambda)  P(B=1|\beta,\lambda) P(\lambda) \,d\lambda
$$

The desired inequality is deduced in the following way. As all probabilities are assumed to satisfy Kolmogorov's axioms, probability takes values between $0$ and $1$. That allows to use inequality (\ref{ineq}) with the association 
$$a= P(A=1|\alpha,\lambda), a'= P(A=1|\alpha',\lambda), b= P(B=1|\beta,\lambda), b'= P(B=1|\beta',\lambda)$$
 to obtain, after multiplying by $P(\lambda)$ and integrating, 
\begin{align}
-1 \leq \int_{\Lambda}   \{ P(A=1|\alpha,\lambda)  P(B=1|\beta,\lambda)  +  P(A=1|\alpha,\lambda)  P(B=1|\beta',\lambda)  +  \nonumber \\ 
  P(A=1|\alpha',\lambda)  P(B=1|\beta,\lambda)   - P(A=1|\alpha',\lambda)  P(B=1|\beta',\lambda) -  \\
  P(A=1|\alpha,\lambda) - P(B=1|\beta,\lambda)  \} P(\lambda) \,d\lambda
  \leq 0 \nonumber
\end{align}

This allows to derive the desired inequality, from now on referred as inequality 1:
\begin{align}
	-1 \leq \nonumber 
P(A=1,B=1|\alpha,\beta)  
&+P(A=1,B=1|\alpha',\beta) \nonumber \\
+P(A=1,B=1|\alpha,\beta') 
&-P(A=1,B=1|\alpha',\beta') \nonumber \\
-P(A=1|\alpha) 
&-P(B=1|\beta)  
\leq 0. 
\label{inequality1}
\end{align}
In brief, this inequality required the following assumptions in its proof: probability space 1, Kolmogorov's axioms, locality and $\lambda$-independence.

\subsection{Bell inequality in probability space 2}

Assuming the existence of the probability space 2, the demonstration of its Bell inequality is straightforward
\cite{eberhard1977bell, redhead1987incompleteness,demuynck1986bell} .

Since $A_{\alpha},A_{\alpha'},B_{\beta},B_{\beta'}$ only assume the values $0$ and $1$, it is possible to use again inequality (\ref{ineq}), making the assignation
$$a= A_{\alpha}, \, a'= A_{\alpha'},\, b= B_{\beta},\, b'= B_{\beta'}$$ 
Multiplying by $P(A_{\alpha},A_{\alpha'},B_{\beta},B_{\beta'},\lambda)$, i.e. the joint probability, and averaging over all the variables it is obtained:
\begin{align}
	-1 \leq \int_{\Lambda} 
	\sum_{A_{\alpha},B_{\beta},A_{\alpha'},B_{\beta'} = \{0,1\}}  
	\{ 
		A_{\alpha}B_{\beta} 
		+A_{\alpha'}B_{\beta} 
		&+A_{\alpha}B_{\beta'} 
		-A_{\alpha'}B_{\beta'}    \\ \nonumber
		-A_{\alpha}
		-&B_{\beta} 
	\} 
	P(A_{\alpha},A_{\alpha'},B_{\beta},B_{\beta'},\lambda) \,d\lambda
	  \leq 0 
	  \label{preInequality2}
\end{align}

In each realization of the Bell experiment, the outputs $A_{\alpha}, A_{\alpha'}, B_{\beta}, B_{\beta'}$ can have only values $0$ or $1$. Therefore, according to section \ref{expectedValues}, it is possible to use identity (\ref{xyproduct}). Each term can be rewritten as
\begin{align}
	\int_{\Lambda} 
		\sum_{A_{\alpha},B_{\beta},A_{\alpha'},B_{\beta'} = \{0,1\}}  
			A_{\alpha}B_{\beta} \quad
		P(A_{\alpha},A_{\alpha'},B_{\beta},B_{\beta'},\lambda)
	 \,d\lambda
	=
	P(A_{\alpha}=1,B_{\beta}=1).
\end{align}
Then, inequality (\ref{preInequality2}) takes the desired form:
\begin{align}
	-1 \leq 
	P(A_{\alpha}=1,B_{\beta}=1) 
	&+P(A_{\alpha}=1,B_{\beta'}=1) \nonumber \\
	+P(A_{\alpha'}=1,B_{\beta}=1) 
	&-P(A_{\alpha'}=1,B_{\beta'}=1) \nonumber \\
	-P(A_{\alpha}=1) 
	&-P(B_{\beta}=1) 
	\leq 0.
\label{inequality2}
\end{align}
From now on this will be called inequality 2.

To summarize, to demonstrate the validity of the inequality 2, Eq. (\ref{inequality2}), probability space 2 and the validity of the Kolmogorov axioms is required. On the other hand, there is no need to invoke locality nor $\lambda$-independence.

\subsection{Bell inequality in a third probability space}

It is possible to derive a third form of the inequalities, one that has the angles and at the same time associates values to all observables. It must be based in a sample space that takes into account all the quantities:

\begin{align}
	(A_{0\degg},A_{45\degg},\alpha,B_{0\degg},B_{45\degg},\beta,\lambda)
\end{align}

Using the biggest event space a new probability space is formed. The inequality, referred as inequality 3, in this space has the form:

\begin{align}
	-1 \leq \nonumber 
P(A_{0\degg}=1,B_{0\degg}=1|\alpha=\theta,\beta=\omega)  
&+P(A_{0\degg}=1,B_{45\degg}=1|\alpha=\theta',\beta=\omega) \nonumber \\
+P(A_{45\degg}=1,B_{0\degg}=1|\alpha=\theta,\beta=\omega') 
&-P(A_{45\degg}=1,B_{45\degg}=1|\alpha=\theta',\beta=\omega') \nonumber \\
-P(A_{0\degg}=1|\alpha=\theta) 
&-P(B_{0\degg}=1|\beta=\omega)  
\leq 0.
\label{inequality3}
\end{align}

The proof can be done using the same argument of inequality 1. It turns out that the analogous inequality in this space also needs the hypotheses of inequality 1\cite{berkovitz2012world}, so the introduction of simultaneous values has not impact in the hypotheses when compared to inequality 1.

Let's summarize these possibilities in a table:

\begin{table}[ht]
	\caption{The hypothesis needed for inequalities 1, 2 and 3}
	\centering
	\begin{tabular}{|c|c|c|c|}
	\hline
	To prove	& Inequality 1  & Inequality 2 & Inequality 3\\
	\hline
		Sample space & $(A,\alpha,B,\beta,\lambda)$ & 
		$(A_{0\degg},A_{45\degg}
		B_{0\degg},B_{45\degg},\lambda)$ &
		 $(A_{0\degg},A_{45\degg},\alpha,B_{0\degg},B_{45\degg},\beta,\lambda)$\\
			& Kolmogorov Axioms & Kolmogorov Axioms & Kolmogorov Axioms\\
			& Locality &   & Locality	\\ 
			& $\lambda$-independence &  & $\lambda$-independence\\
			\hline
	\end{tabular}
	\label{shortTable}
	\end{table}


\subsection{Bell inequalities without probability}

It is possible to derive inequality 2 without making any explicit reference to probabilities, only to relative frequencies \cite{eberhard1977bell, redhead1987incompleteness}.

In an experiment repeated $N$ times, $A_{\alpha}^i$ is meant to represent the value of the variable $A_{\alpha}$ in the $i$-th repetition of the experiment, and the same notation for the other variables.
Since each variable only assumes the values $0$ or $1$, it is possible again to use (\ref{ineq}) with the association:
$$a= A_{\alpha}^i,\, a'= A_{\alpha'}^i,\, b= B_{\beta}^i,\, b'= B_{\beta'}^i$$ 
Summing over all the repetitions of the experiment, and dividing by $N$, we have
\begin{align}
-1 \leq  
		\frac{1}{N} \sum_i A_{\alpha}^i B_{\beta}^i  +  
		\frac{1}{N} \sum_i A_{\alpha}^i B_{\beta'}^i  +  
		\frac{1}{N} \sum_i A_{\alpha'}^i  B_{\beta}^i  - 
		\frac{1}{N} \sum_i A_{\alpha'}^i  B_{\beta'}^i  -   
		\frac{1}{N} \sum_i A_{\alpha}^i  -  
		\frac{1}{N} \sum_i B_{\beta}^i
\leq 0.
\end{align}

The hypothesis are:
\begin{itemize}
\item The outputs $A_{\alpha}, A_{\alpha'}, B_{\beta}, B_{\beta'}$ have assigned values in all the runs, although only two of them are measured in each run.
\item The observed product averages $\sum_i A_{\alpha}^i B_{\beta}^i $ can be experimentally determined only for the subset of the results in which the angles $\alpha, \beta$ were selected, while the theoretical average refers to the whole set of values. A version of the {\em fair sampling} assumption is required, to associate the observed and predicted values \cite{khrennikov2009}.
\end{itemize}

\section{Interpretations of inequality violation}

Violations of the Bell inequalities have been observed since the first experiments \cite{aspect1976proposed, clauser1969proposed}. The original setting raised a list of objections regarding the adequacy of the experimental setup, for instance efficiency of detectors, space time separation of the detectors on both sides of experiment, etc.\cite{branciard2011detection, barrett2002quantum, larsson2014loopholes, valdenebro2002assumptions}. Most of the original objections have been answered with more refined settings, closing at the same time many of the loopholes discussed in the literature \cite{tittel1998violation, handsteiner2017cosmic, rowe2001experimental, hensen2015loopholefree,giustina2013bell}. Despite of the remarkable experimental creativity and amazing technological developments, the debate about the validity and the interpretations of the observed violation of Bell inequalities remains vivid. In what follows we analyze the consequences, assuming that the experiments actually show a violation of Bell's inequalities.

In the above discussion it has been established that there are different inequalities with different hypotheses, listed in Table II. For each one of them, accepting the experimental violation implies that some of the hypotheses must be questioned.

The inequalities 1 and 3 are based in the same set of hypothesis. Most of the literature around this subject concentrates on locality and $\lambda$-independence. Locality, as its name suggests, is interpreted trough its relation with space, and refers to the impossibility of mutual influence between events in space-time whose interval is space-like. The probability independence, or statistical independence, is interpreted as independence of the physical situation, and is sometimes related with contextuality \cite{hall2016significance,khrennikov2021can}. Their consequences have been widely discussed in many excellent texts \cite{bell2004theory,eberhard1978bell, eberhard1978bell, fine1980correlations, jarrett1984physical, redhead1987incompleteness, berkovitz1998aspects,maudlin2014what,khrennikov2020two}. We will not abound on them.

The violation of any of the three inequalities can be interpreted, alternatively, questioning any element of the probability space, i.e. the sample space, the event space or the probability measure. 

First, consider that the failure is in the sample space. That is, the considered sample space does not represent correctly the experimental situation. 
Since all sample spaces have hidden variables included,
it could be tempting to conclude that the mere existence of hidden variables must be rejected, invalidating the use of any of the sample spaces. This position is also close to the orthodox quantum mechanics. In particular, it is related with the debate of completeness of quantum mechanics. However, it is difficult to hold this position, since there are theories like bohmian mechanics that use {\it non-local} hidden variables and successfully reproduce the results of quantum experiments \cite{bohm1952suggested,bohm1952reply}.

An argument can be stated against the use of the sample spaces 2 and 3. From the old times of quantum mechanics, the simultaneous assignment of values for non-commuting observables has been rejected \cite{murdoch1987niels}, and these sample spaces do such value assignment. 
In the same line, an instrumentalist position will argue against these spaces, since there are values that cannot be simultaneously observed, condensed in the famous dictum {\em unperformed experiments have no results} \cite{peres1978unperformed,berkovitz2012world}. 

To continue, consider that the failure is in the event space. It must be closed under unions and intersections.

It is possible to accept the sample spaces 2 and 3, but reject their event space, arguing that only events that can be observed are to be considered. 
From this point of view, the event $\{A_{0\degg}=1,A_{45\degg}=1\}$ must be rejected since it cannot be observed. This is a problem since $\{A_{0\degg}=1\}$ and $\{A_{45\degg}=1\}$ can be observed, so its intersection should be an event.
This is enough to invalidate the event space.

Continuing with the elements to be questioned, the probability measure is the next in the list. It is required to assign probability to all events and to fulfill Kolmogorov's axioms. Therefore, it can be argued that the sample and event spaces do apply for the situation, but not all events have an assigned probability. 
The existence of the joint probability for the four variables of interest, i.e. $P( A_{\alpha}, A_{\alpha'}, B_{\beta}, B_{\beta'})$, is a necessary and sufficient condition to deduce Bell inequality 2. Therefore the inequality violation can be interpreted as a non existent joint probability. This is argument is due to Fine \cite{fine1982hidden}, and has been explored in \cite{fine1982joint, muller2001minimalist, demuynck1986bell,demuynck1984alternative,khrennikov2007bell}. 

The interpretation of violations of Bell's inequalities as invalidating the sample space, the event space and/or the probability measure, can be be summarized as ``Bell inequality violation just proves that there cannot be a reduction to one common probability space.'' \cite{nieuwenhuizen2011contextuality,delapena1972hiddenvariable,khrennikov2007bell,khrennikov2021can,pitowsky1982resolution}.

Finally, the last hypothesis to be questioned is the applicability of the Kolmogorov axioms. Another way to understand the violation of Bell inequalities is to accept the validity of the probability space, and question the applicability of the Kolmogorov axioms (\ref{kolmogorovAxioms}) to this situation. The axioms guarantee that probabilities have values between zero and one, sum up to one, etc. 

In the context of quantum mechanics, a phase space including non-commuting observables like position and momentum is widely employed as probability space, but the "probabilities" defined there do not fulfill the non-negativity Kolmogorov's postulate, like the Wigner quasi-probability distribution. There are also distributions which are non-negative, but their marginals do not coincide with the quantum mechanical probabilities in position or momentum, like the Husimi distribution \cite{wigner1932quantum,spekkens2008negativity}. There have been also works relating contextuality and negative probabilities in areas outside of physics \cite{debarros2016negative}.

To sumarize: When the Bell's inequalities are formulated in the probability spaces 2 and 3, their violation can be justified denying the pertinence of these sample spaces, or the event spaces, as being not closed under unions and intersections. It is also possible to point out some events that do not have an associated probability, or to exhibit probabilities that do not fulfill the property of being between one and zero.
Inequality 1 is not affected by many of these concerns, which are based in the counterfactual character of elements in the sample spaces 2 and 3. Sample space 1 does not assume the existence of simultaneous values of incompatible outputs in each experimental run. For this reason, if the probability space 1 is employed to deduce the inequality 1, and the Kolmogorov framework is assumed, it is somehow natural to conclude that either locality or $\lambda$-independence must be questioned to explain the inequality violation.

A final remark refers to the Kolmogorov axioms as the best theoretical representation of probability theory. It is worth to remember that the origin of probability can be traced back to the XVII century. Extensive and valuable studies about  the nature of probability were written \cite{gillies2000philosophical, hacking2006emergence, daston1988classical} before the Kolmogorov axioms appeared in 1932 \cite{kolmogorov2018foundations}. Each school of thought has important differences regarding event spaces and probability measures. Although all schools coincide in Kolmogorov axioms for textbook cases, the Bell scenario appear as the perfect situation to raise the differences \cite{svetlichny1988bell, berkovitz2012world,khrennikov2002frequency}. 
 
\section{Conclusion}

Bell type experiments can be analyzed employing different probability spaces under Kolmogorov's axioms. Each of them are based in some hypothesis, listed in table \ref{shortTable}. Under some simple mathematical manipulations they lead to different versions of the Bell-CHSH inequalities, all of them mathematically sound but not equivalent. Their observed violation implies that at least one of the hypothesis must be refuted.  
Employing the probability space 1 and the validity of the Kolmogorov axioms lead many authors to conclude that any hidden variable model must be non-local and/or contextual. On the other hand, probability space 2 offers other interpretations: that values cannot be simultaneously assigned to all variables, or that the values can be assigned but joint probabilities cannot be properly defined, or that such situations pertain to different probability spaces. 

The observed violation of the Bell-CHSH inequalities impose conditions on any physical theory aimed to describe these observations. The theoretical or philosophical frame where the description is performed depends on the moment and the interest of each community. Those elements are at the base of each interpretation. Particular probability theories, like frequentism,  of philosophical views like realism, would add elements that may dismiss one or more probability spaces, narrowing the set of hypothesis and changing the epistemological interpretation.

The intention of this contribution is to emphasize the richness of the possibilities and that, in the present use of probability, there is no evidence that one option should be selected over the others. The same experimental results lead to different conclusions depending on the description used.  

As different probability theories have different interpretations on what a probability refers to, they have important implications in the analysis of the violation of the Bell-CHSH inequalities which we plan to analyze in a future work.

\authorcontributions{AFLS-L made the investigation and wrote the original draft, MG and JGH contributed with the conceptualization and writing, review and editing.}

\funding{This research was partially funded by DGAPA-UNAM project IN104020.}

\conflictsofinterest{The authors declare no conflict of interest.} 

\appendixtitles{yes} 
\appendixstart
\appendix
\section{Derivation of the numerical inequality}

Let be 4 real variables $x,x',y,y' \in [-1,1]$. The function 
$$
f(x,x',y,y') \equiv x y + x y' + x' y - x' y' = x (y + y') + x' (y- y') = y ( x + x') + y' (x - x')
$$
has critical points defined by
$$
\frac{\partial f}{\partial x} = y + y' = 0 , \,\,\,\,
\frac{\partial f}{\partial x'} = y - y' = 0 , \,\,\,\,
\frac{\partial f}{\partial y} = x + x' = 0 , \,\,\,\,
\frac{\partial f}{\partial y'} = x - x' = 0.
$$
The above conditions imposse $x = \pm x', \, y = \pm y'$.
Given that the extreme values each variable can have are $\pm 1$, the function $f$ in this domain is bounded as 
$$
-2 \leq f(x,x',y,y') \leq 2.
$$
  
Under the linear transformation $x = 2 a -1, \, x' = 2 a' -1, \, y = 2 b -1, y' = 2 b' -1 $, the new variables have the domain $a,a',b,b' \in [0,1]$, the function $f$ becomes
$$
f(a,a',b,b') \equiv a b + a b' + a' b - a' b' - a - b,
$$
and the inequality reads
$$
-1 \leq  a b + a b' + a' b - a' b' - a - b \leq 0.
$$

\end{paracol}
\reftitle{References}



\end{document}